\documentstyle[12pt]{article}
\newcommand{\beq}{\begin{equation}}
\newcommand{\eeq}{\end{equation}}
\newcommand{\beqa}{\begin{eqnarray}}
\newcommand{\eeqa}{\end{eqnarray}}
\newcommand{\beqan}{\begin{eqnarray*}}
\newcommand{\eeqan}{\end{eqnarray*}}
\newcommand{\bdes}{\begin{description}}
\newcommand{\edes}{\end{description}}
\newcommand{\ben}{\begin{enumerate}}
\newcommand{\een}{\end{enumerate}}
\newcommand{\bfl}{\begin{flushleft}}
\newcommand{\efl}{\end{flushleft}}
\newcommand{\ba}{\begin{array}}
\newcommand{\ea}{\end{array}}
\newcommand{\btab}{\begin{tabular}}
\newcommand{\etab}{\end{tabular}}
\newcommand{\bit}{\begin{itemize}}
\newcommand{\eit}{\end{itemize}}
\newcommand{\no}{\nonumber}

\newcommand{\ol}{\overline}
\newcommand{\ra}{\rightarrow}

\newcommand{\ve}{\varepsilon}
\newcommand{\vp}{\varphi}

\newcommand{\dg}{\dagger}

\newcommand{\del}{\partial}

\newcommand{\cL}{{\cal L}}

\newcommand{\toG}{\stackrel{G}{\to}}
\normalsize
\begin{document}
\begin{titlepage}
\begin{flushright}
UWThPh-1994-24\\
July 1994\\
hep-ph/9407240
\end{flushright}
\vspace*{2cm}
\begin{center}
{\Large \bf The Pion--Nucleon Interaction as an\\[5pt]
Effective Field Theory*}\\[50pt]
G. Ecker  \\
Institut f\"ur Theoretische Physik \\
Universit\"at Wien \\
Boltzmanngasse 5, A-1090 Vienna, Austria
\vfill
{\bf Abstract} \\
\end{center}
\noindent
After a brief survey of effective field theories,
the linear $\sigma$ model is discussed as a prototype of an
effective field theory of the standard model below the
chiral--symmetry--breaking scale. Although it can serve as a toy
model for the pion--nucleon interaction, the linear $\sigma$ model is
not a realistic alternative to chiral perturbation theory.
The heavy--baryon approach to chiral perturbation theory allows for
a systematic low--energy expansion of Green functions and
amplitudes with baryons. The chiral--invariant renormalization of
the effective field theory for the pion--nucleon
system to $O(p^3)$ in the chiral expansion is reviewed.
\vfill
\begin{center}
To appear in the Proceedings of the \\
First International Symposium on Symmetries in Subatomic Physics\\
Taipei, Taiwan, May 16 -- 18, 1994
\end{center}
\vfill
\noindent * Work supported in part by FWF, Project No. P09505--PHY and
by HCM, EEC--Contract No. CHRX--CT920026 (EURODA$\Phi$NE)
\end{titlepage}

\renewcommand{\theequation}{\arabic{section}.\arabic{equation}}
\setcounter{equation}{0}
\section{Effective Field Theories}
\label{sec:EFT}
Effective field theories (EFT) are the quantum field theoretical
implementation of the quantum ladder. As the energy increases and smaller
distances are probed, new degrees of freedom become relevant
that must be included in the theory. At the same time,
other fields may lose their status of fundamental fields as the
corresponding states are recognized as bound states of the new degrees
of freedom. An EFT is characterized by a set of
``asymptotic" fields and an energy scale $\Lambda$ beyond which
it must be replaced by a more ``fundamental" theory. The development
of modern physics can be viewed as a sequence of EFT
culminating in the standard model of the fundamental interactions.
We have every reason to believe that we have not arrived at the
final stage of this development yet. All we know for sure, however, is
that the associated energy scale $\Lambda$ of the standard model is
bigger than $O(100\ {\rm GeV})$.

Conversely, as the energy is lowered, some degrees of freedom are
frozen out and disappear from the accessible spectrum of states.
To model the EFT at low energies, we rely especially
on the symmetries of the ``fundamental" underlying theory, in addition
to the usual axioms of quantum field theory embodied in a corresponding
effective Lagrangian. An important feature of an EFT
is that its Lagrangian must contain {\bf all} terms
allowed by the symmetries of the fundamental theory
for the given set of fields \cite{Wein79}.
This completeness guarantees that the EFT
is indeed the low--energy limit of the fundamental theory.

Two types of EFT can be distinguished \cite{Prag}.
\subsection*{\bf A. Decoupling EFT}
For energies below the scale $\Lambda$, all heavy (with respect
to $\Lambda$) degrees of freedom are integrated out leaving only
the light degrees of freedom in the effective theory. No light
particles are generated in the transition from the fundamental
to the effective level. The effective Lagrangian has the general
form
\beq
\cL_{\rm eff} = \cL_{d\leq 4} + \sum_{d>4} \frac{1}{\Lambda^{d-4}}
\sum_{i_d} g_{i_d} O_{i_d}  \label{eft}
\eeq
where $\cL_{d\leq 4}$ contains the potentially renormalizable terms
with operator dimension $d \leq 4$, the $g_{i_d}$ are dimensionless
coupling constants expected to be of $O(1)$,
and the $O_{i_d}$ are monomials
in the light fields with operator dimension $d$. At energies
much below $\Lambda$,
corrections due to the non--renormalizable parts ($d>4$) are suppressed
by powers of $E/\Lambda$. In such cases, $\cL_{d\leq 4}$ can be regarded
as the ``fundamental'' Lagrangian at low energies.

\vspace*{.3cm}
\noindent
Examples of decoupling EFT:
\bdes
\item[i.] {QED for $E \ll m_e$} \\
For energies much smaller than the electron mass, the electrons
are integrated out to yield the Euler--Heisenberg Lagrangian
for light-by-light scattering \cite{EuHe}.
\item[ii.] {Weak interactions for $E \ll M_W$} \\
At low energies, the weak interactions reduce to the Fermi
theory with $d=6$.
\item[iii.] {The standard model for $E \ll 1 \mbox{ TeV}$} \\
There are many candidates for an underlying theory at smaller distances
(composite Higgs, SUSY, grand unification, superstrings,\dots).
With the exception of the Higgs sector, the standard model does
not provide any clues for the scale $\Lambda$. There is no
experimental evidence for terms in the effective Lagrangian with
$d > 4$.
\edes
\subsection*{\bf B. Non--decoupling EFT}
The transition from the fundamental to the effective level occurs through a
phase transition via the
spontaneous breakdown of a symmetry generating light
($M \ll \Lambda$) pseudo--Goldstone bosons. Since a spontaneously broken
symmetry relates processes with different numbers of Goldstone bosons,
the distinction between renormalizable ($d\leq 4$) and non--renormalizable
($d>4$) parts in the effective Lagrangian like in (\ref{eft})
becomes meaningless. The effective Lagrangian of a non--decoupling EFT is
intrinsically non--renormalizable. Nevertheless, such
Lagrangians define perfectly consistent quantum field theories
\cite{Wein79,GL1,GL2,HL93}. Instead of the operator dimension as
in (\ref{eft}), the number of derivatives of the fields distinguishes
successive terms in the Lagrangian.

The general structure of effective Lagrangians with spontaneously broken
symmetries is largely independent of the specific physical
realization. This is exemplified by two examples in particle physics.
\bdes
\item[a.] {The standard model without Higgs bosons} \\
Even if there is no explicit Higgs boson, the gauge symmetry $SU(2)\times U(1)$
can be spontaneously broken to $U(1)_{\rm em}$ (heavy Higgs scenario).
As a manifestation of the universality of Goldstone
boson interactions, the scattering of longitudinal gauge vector bosons
is in first approximation analogous to $\pi\pi$ scattering.
\item[b.] {The standard model for $E< 1 \mbox{ GeV}$} \\
At low energies, the relevant degrees of freedom of the standard
model are not quarks and gluons, but the pseudoscalar mesons
and other hadrons. The pseudoscalar mesons play a special role as
the pseudo--Goldstone bosons of spontaneously broken chiral
symmetry. The standard model in the hadronic sector
at low energies is described by a non--decoupling EFT  called chiral
perturbation theory (CHPT).
\edes

\renewcommand{\theequation}{\arabic{section}.\arabic{equation}}
\setcounter{equation}{0}
\section{From the Linear $\sigma$ Model to CHPT}
The linear $\sigma$ model \cite{SGML} is
a seeming counterexample to the classification of Sect. \ref{sec:EFT}:
it is a renormalizable quantum field
theory describing the spontaneous breaking of chiral symmetry.
It is instructive to rewrite it in the form of a non--decoupling
EFT to demonstrate the price of renormalizability: although
it has the right symmetries by construction, the linear $\sigma$ model
is not general
enough to describe the real world \cite{GL1}. It is instructive
as a toy model, but it should not be mistaken for the EFT
of QCD at low energies.

We rewrite the $\sigma$ model Lagrangian for the pion--nucleon system
\beq
\cL_\sigma = {1\over 2} \left(\del_\mu\sigma\del^\mu\sigma +
 \del_\mu\vec{\pi}\del^\mu\vec{\pi}\right) -{\lambda\over 4}
\left(\sigma^2 + \vec{\pi}^2 - v^2\right)^2 + \ol\psi
 ~i\not\!\partial\psi - g \ol\psi \left(\sigma + i\vec{\tau}\vec{\pi}
\gamma_5\right)\psi \label{eq:Lsig1}
\eeq
$$
\psi = { p \choose n}
$$
in the form
\beq
\cL_\sigma = {1\over 4}\langle \del_\mu\Sigma\del^\mu\Sigma \rangle
- {\lambda\over 16} \left( \langle \Sigma^\dg\Sigma \rangle - 2 v^2\right)
^2 + \ol{\psi_L} ~i\not\!\partial\psi_L + \ol{\psi_R} ~i\not\!\partial\psi_R
- g \ol{\psi_R} \Sigma\psi_L - g \ol{\psi_L} \Sigma^\dg\psi_R
\label{eq:Lsig2} \eeq
$$
\Sigma = \sigma{\bf 1} - i\vec{\tau}\vec{\pi} ~,\quad\qquad
\langle A \rangle = tr A
$$
to exhibit the chiral symmetry $G=SU(2)_L\times SU(2)_R$~:
$$
\psi_A \toG g_A \psi_A~,\qquad g_A \in SU(2)_A \quad(A=L,R)~,
\qquad \Sigma \toG g_R\Sigma g_L^\dg~.
$$
For $v^2>0$, the chiral symmetry is spontaneously broken and the
``physical" fields are the massive field $\hat\sigma = \sigma - v$
and the Goldstone bosons $\vec \pi$. However, the Lagrangian
with its non--derivative couplings for the $\vec\pi$ seems to be
at variance with the Goldstone theorem predicting a vanishing amplitude
whenever the momentum of a Goldstone boson goes to zero.

In order to make the Goldstone theorem manifest in the Lagrangian,
we perform a field transformation from the original fields
$\psi$, $\sigma$, $\vec\pi$ to a new set $\Psi$, $S$, $\vec\vp$
through a polar decomposition of the matrix field $\Sigma$:
\beq
\Sigma = (v+S)U(\vp) ~,\quad\qquad \Psi_L=u\psi_L ~,\qquad
\Psi_R=u^\dg \psi_R \label{eq:ftrans}
\eeq
$$
S^\dg = S ~,\qquad U^\dg=U^{-1} ~,\qquad \det U=1 ~,\qquad U=u^2~.
$$
Under a chiral transformation,
\beq
u(\vp) \toG g_R u(\vp) h(g,\vp)^{-1}
= h(g,\vp) u(\vp) g_L^{-1}~,\quad\qquad g=(g_L,g_R)\in G \label{eq:nlin}
\eeq
defines a non--linear realization of $G$ via the compensator
field $h(g,\vp)$ \cite{CCWZ}. Consequently,
\beq
U\to g_R U g_L^{-1}~, \qquad S\to S~, \qquad \Psi_A\to h(g,\vp)\Psi_A
\quad (A=L,R)~.
\eeq
In the new fields, the $\sigma$--model Lagrangian (\ref{eq:Lsig1})
takes the form
\beqa
\cL &=& {v^2\over 4}(1 + {S\over v})^2\langle u_\mu u^\mu \rangle\no\\
&+& \ol\Psi ~i \not\!\nabla \Psi + {1\over 2} \ol\Psi
\not\!u \gamma_5\Psi - g(v+S)\ol\Psi \Psi + \ldots \label{eq:Lsig3}
\eeqa
with a covariant derivative $\nabla = \del + \Gamma$ and
\beqa
u_\mu(\vp) &=& i (u^\dg \partial_\mu u - u \partial_\mu u^\dg)\no\\
\Gamma_\mu(\vp) &=& \frac{1}{2} (u^\dg \partial_\mu u
+ u \partial_\mu u^\dg) ~.
\label{eq:Gu}
\eeqa
The kinetic term and the self--couplings of the scalar field $S$
are omitted in (\ref{eq:Lsig3}).

We can draw the following conclusions:
\bdes
\item[i.]
The Goldstone theorem is now manifest at the Lagrangian level: the
Goldstone bosons $\vec\vp$ contained in the matrix fields $u_\mu(\vp)$,
$\Gamma_\mu(\vp)$ have derivative couplings only.
\item[ii.]
By construction, S--matrix elements are unchanged under the field
transformation (\ref{eq:ftrans}), but the Green functions are very
different. For instance, in the pseudoscalar meson sector the field $S$
does not contribute at all
at lowest order, $O(p^2)$, whereas $\hat\sigma$ exchange is essential
to repair the damage done by the non--derivative couplings of the
$\vec\pi$. The linear $\sigma$ model ascribes an importance to the field
$\hat\sigma$ which does not match the relevance of the
corresponding physical state. At $O(p^4)$ in the meson sector,
the chiral Lagrangian is known to be dominated by meson--resonance
exchange \cite{GL1,EGPR}. However, the scalar resonances are much less
important than the vector and axial--vector mesons \cite{EGPR}.
\item[iii.]
The manifest renormalizability of the Lagrangian (\ref{eq:Lsig1}) has
been traded for the manifest chiral structure of (\ref{eq:Lsig3}).
Of course, the Lagrangian (\ref{eq:Lsig3}) is still renormalizable,
but this renormalizability has its price. It requires specific
relations between various couplings that have nothing to do with
chiral symmetry and, which is worse, are not in agreement with
experiment. For instance, the model contains the Goldberger--Treiman
relation \cite{GoTr} in the form ($m$ is the nucleon mass)
\beq
m = g v \equiv g_{\pi NN} F_\pi~.
\eeq
Thus, instead of the physical value $g_A=1.26$ for the axial--vector
coupling constant $g_A$ the model has $g_A=1$ (compare with the CHPT Lagrangian
(\ref{eq:piN1}) below). As already emphasized,
the problems with the linear $\sigma$ model are even more severe
in the meson sector. In the form of (\ref{eq:Lsig1})
or (\ref{eq:Lsig3}), the linear $\sigma$ model is a toy model, but
not a realistic EFT of QCD in the pion--nucleon sector.
\edes

Of course, nobody uses the original $\sigma$ model (\ref{eq:Lsig1})
nowadays for actual phenomenological analysis. By introducing additional
terms in the Lagrangian, one may reconcile the model with experimental
data for the price of abandoning renormalizability. Compared with the
alternative general approach of CHPT described in the next section,
such a procedure has several
conceptual drawbacks that tend to obscure the relation to the underlying
``fundamental" theory of QCD. To make the point, let me consider
an example of such an approach inspired by the linear $\sigma$ model.

In the model of Goudsmit et al. \cite{Goud}, the relevant interaction
terms (among others, including vector mesons) are given by the Lagrangians
\beqa
\cL_{\pi N}^{\rm int} &=&
-{g_{\pi NN}\over 1+x}~\ol\psi \gamma_5\vec\tau\left(
i x \vec\pi + {1\over 2 m_N} \not\!\partial\vec\pi \right)\psi\no\\
\cL_\sigma^{\rm int} &=& - g_{\pi\pi\sigma} M_\pi \vec\pi\vec\pi\sigma-
g_{\sigma NN} \ol\psi \sigma\psi~.
\eeqa
In the last paper of Ref.~\cite{Goud}, the authors have performed
a fit of their model to the $\pi N$ phase shifts at low energies.
As explicitly stated in their paper, chiral symmetry is nowhere
implemented. One interesting consequence of their analysis is that
the pseudoscalar mixing parameter $x$ is strongly correlated
with the effective scalar coupling $G_\sigma$ defined as
$$
G_\sigma = {g_{\pi\pi\sigma}g_{\sigma NN}\over M^2_\sigma}
$$
and with a corresponding vector coupling $G^V_\rho$. Two extreme cases
are listed in Table \ref{tab:Goud}. Although the authors deplore
that the parameters $x$, $G_\sigma$ and $G^V_\rho$ cannot be well
determined at present, the content of Table \ref{tab:Goud} is actually
a field transformation in action. For $x=0.2$, the scalar parameter
$G_\sigma$ must be big in order to repair the chiral--symmetry violation
by the pseudoscalar
pion--nucleon coupling. For a pure pseudo--vector coupling ($x=0$),
$G_\sigma$ is significantly smaller while the vector
exchange becomes more important. The quality of fit is the same
in both cases as one would expect for identical $S$--matrix elements.

\begin{table}
\caption{Parameters of the model of Ref.~\protect\cite{Goud} extracted
from a fit to $\pi N$ phase shifts. The quality of fit is the same
for the two sets of parameters.}\label{tab:Goud}
$$
\begin{tabular}{|clc|ccc|ccc|} \hline
 & $x$&  & & $G_\sigma$ (GeV$^{-2}$) &  & & $G^V_\rho$ (GeV$^{-2}$)&
\\ \hline
 & 0.2&  & & 43         &  & & 30                     &  \\
 & 0  &  & & 23         &  & & 60                     &  \\ \hline
\end{tabular}
$$
\end{table}

What can one learn from an analysis of this type? The suggested
greater generality of the model compared to a pure pseudo--vector
pion--nucleon coupling is spurious.
In particular, there is nothing in the $\pi N$ phase shifts that
would argue against using
a manifestly chirally symmetric framework like CHPT with $x=0$.
At the same time, the scalar field is reduced to a role in agreement
with the status of the corresponding $I=0$ $s$--wave meson resonance
$f(1000)$ in the 1994 edition of the Review of Particle Properties
\cite{Mont}:
an inconspicuous, highly elastic and very broad ($\Gamma\simeq 700
\mbox{ MeV}$) $\pi\pi$ resonance. There is nothing special about
the $\sigma$ meson \cite{MeiComm}~!

\renewcommand{\theequation}{\arabic{section}.\arabic{equation}}
\setcounter{equation}{0}
\section{Heavy mass expansion}
\label{sec:HME}
CHPT for the $\pi N$ system starts from the most general chiral--invariant
effective Lagrangian \cite{GL1,GSS}
\beqa
\cL_{\rm eff} &=& \cL_{\rm meson} + \cL_{\pi N}  \label{eq:Leff}\\
\cL_{\rm meson} &=& \cL_2 + \cL_4 + \ldots  \no\\
\cL_2 &=& \frac{F^2}{4} \langle u_\mu u^\mu + \chi_+ \rangle
\label{eq:L2} \\
\cL_{\pi N} &=& \cL_{\pi N}^{(1)} + \cL_{\pi N}^{(2)} + \cL_{\pi N}^{(3)}
+ \ldots  \no\\
\cL_{\pi N}^{(1)} &=& \ol \Psi (i \not\!\nabla - m + \frac{g_A}{2}
\not\!u \gamma_5) \Psi \label{eq:piN1}
\eeqa
$$
u_\mu = i \{ u^\dg (\partial_\mu - i r_\mu)u -
u (\partial_\mu - i l_\mu) u^\dg\} ~, \qquad
\chi_+ = u^\dg \chi u^\dg + u \chi^\dg u ~,
$$
$$
\nabla = \del + \Gamma ~,\qquad
\Gamma_\mu = \frac{1}{2} \{ u^\dg (\partial_\mu - i r_\mu)u +
u (\partial_\mu - i l_\mu)u^\dg\} ~.
$$
The fields $u_\mu$, $\Gamma_\mu$ are as in (\ref{eq:Gu}),
except that they now contain also the external gauge fields (like
photon and $W$ boson)
\beq
r_\mu  =  v_\mu + a_\mu ~,\quad\qquad
l_\mu  =  v_\mu - a_\mu ~.
\eeq
The scalar field $\chi$ includes the quark masses responsible
for the explicit chiral--symmetry breaking:
$$
\chi = 2 B_0 ~\mbox{diag}(m_u,m_d,m_s)+\dots
$$
At lowest order in the derivative expansion, the effective chiral
Lagrangian for the
pion--nucleon system contains four parameters, the nucleon mass
$m$, the axial coupling constant $g_A$ and the mesonic parameters
$F$ and $B_0$ related to the pion decay constant and the quark condensate,
respectively:
\beqa
F_\pi &=&  F[1 + O(m_{quark})] = 93.2 MeV \no\\
\langle 0|\bar u u |0\rangle &=& - F^2 B_0[1 + O(m_{quark})]\\
M^2_{\pi^+} &=& B_0 (m_u + m_d)[1+O(m_{quark})]~.\no
\eeqa
Comparing with (\ref{eq:Lsig3}), we realize that the $\sigma$ model
has indeed the correct chiral structure, but with the axial coupling
constant $g_A=1$.

The effective Lagrangian (\ref{eq:Leff}) is the starting point for a systematic
low--energy expansion of Green functions and amplitudes. To satisfy
unitarity and analyticity, it is essential to consider the
chiral Lagrangian not only at tree level, but to take it seriously
as an EFT by including loops. There
is however a difference between the purely mesonic and
the pion--nucleon sector \cite{GSS}: in contrast to mesonic amplitudes,
the loop expansion and the derivative expansion of amplitudes do
not coincide in the presence of baryons. The reason is very simple:
unlike the pseudoscalar
meson masses, the nucleon mass $m$ does {\bf not} vanish in the chiral
limit. Therefore, the nucleon four--momentum can never be soft.
The size of the nucleon mass suggests a simultaneous expansion in
$$
{p\over 4\pi F} \qquad\quad {\rm\bf and} \qquad\quad {p\over m}~,
$$
but there is an essential difference between the two denominators ($p$
stands for a generic meson four--momentum or nucleon three--momentum):
$F$ appears only in vertices
of the effective Lagrangian (\ref{eq:Leff}), while the nucleon mass
enters via the nucleon propagator. To put these two quantities
on the same footing, one has to find a way to move the nucleon
mass from the propagator to the vertices of some effective Lagrangian.

With inspiration from heavy quark effective theory, Jenkins
and Manohar \cite{JM1} have reformulated baryon CHPT in precisely such
a way as to transfer the nucleon mass from propagators to
vertices. The method is called ``heavy baryon CHPT" and it can be
interpreted \cite{MRR} as a clever choice of variables for performing the
fermionic integration in the path integral representation of the
generating functional of Green functions
\beq
e^{iZ[j,\eta,\ol\eta]} = N \int [du d\Psi d\ol\Psi]
\exp\left[ i \left\{ S_{\rm meson} + S_{\pi N} + \int d^4x(\ol\eta \Psi +
\ol\Psi \eta)\right\}\right]~. \label{eq:ZMB}
\eeq
The action $S_{\rm meson}$ $+ S_{\pi N}$ corresponds to the effective
Lagrangian (\ref{eq:Leff}), the external fields $v_\mu$, $a_\mu$, $\chi$
are denoted collectively as $j$ and $\eta,\ol\eta$ are fermionic sources.
Heavy baryon CHPT can be formulated in a manifestly Lorentz covariant way
by defining velocity--dependent fields \cite{Georgi}
\beqa
N_v(x) &=& \exp[i m v\cdot x] P_v^+ \Psi(x) \label{eq:vdf} \\
H_v(x) &=& \exp[i m v\cdot x] P_v^- \Psi(x) \no \\
P_v^\pm &=& \frac{1}{2} (1 \pm \not\!v)~, \qquad v^2 = 1~. \no
\eeqa
In the nucleon rest frame $v = (1,0,0,0)$ and $N_v$, $H_v$ correspond to the
usual non--relativistic projections of a Dirac spinor into upper--
and lower--component Pauli spinors. In general, we may call the $N_v$
($H_v$) the light (heavy) components of the nucleon field $\Psi$.
In the functional integral (\ref{eq:ZMB}), one first integrates out the
heavy components $H_v$ and then expands the resulting non--local
action in a power series in $1/m$. The resulting effective pion--nucleon
Lagrangian contains only the light components $N_v$ together
with the pion fields:
\beq
\cL_{\pi N}(N_v,\vp) = \ol{N_v}(iv \cdot \nabla + g_A S \cdot u)N_v
+ O(p^n)~,\qquad n\geq 2 \label{eq:light}
\eeq
$$
S^\mu = \frac{i}{2} \gamma_5 \sigma^{\mu\nu} v_\nu , \qquad
S \cdot v = 0, \qquad S^2 = - \frac{3}{4} {\bf 1}~.
$$
The nucleon mass appears only in powers of $1/m$ in the higher--order
terms in (\ref{eq:light}). The $N_v$ propagator is
\beq
\frac{i P_v^+}{v \cdot k + i\ve} \label{Nprop}
\eeq
according to the Lagrangian (\ref{eq:light}), independent of the
nucleon mass. Thus, the goal has been achieved to move the
nucleon mass from the propagator to the vertices of an effective
Lagrangian. Consequently, loop and derivative expansion coincide
again as in the mesonic case.

There is a small price one has to pay for the systematic low--energy
expansion in the presence of baryons. Any given order in the
chiral expansion of the generating functional will in general not
be independent of the time--like unit vector $v$, because a
change in $v$ involves different chiral orders (reparametrization
invariance \cite{LuMa}).

\renewcommand{\theequation}{\arabic{section}.\arabic{equation}}
\setcounter{equation}{0}
\section{Renormalization}
\label{sec:renorm}
With the effective pion--nucleon Lagrangian of the last
section, all Green functions and amplitudes with a single incoming
and outgoing nucleon can be calculated in a systematic chiral expansion
\cite{GSS,JM2,RMei,Rho}: nucleon form factors,
$\pi N \ra \pi \ldots \pi N$, $\gamma^* N \ra \pi \ldots \pi N$,
$W^* N \ra \pi \ldots \pi N$.

Up to and including $O(p^2)$, only tree--level amplitudes contribute.
At $O(p^3)$, loop diagrams of the type shown in Figs. 1,2 must be taken
into account. Those diagrams are in general divergent requiring
regularization and renormalization. Since we have a non--decoupling
EFT that is intrinsically non--renormalizable, the divergences must
be cancelled by counterterms of $O(p^3)$. Those counterterms are
part of the general chiral--invariant pion--nucleon Lagrangian
(\ref{eq:light}).

The divergent part of the one--loop functional of $O(p^3)$ can be
calculated in closed form \cite{Ecker} by using the heat kernel method
(see Ref.~\cite{Ball} for a review) for the meson and nucleon propagators
in the presence of external fields (the propagators appearing in Fig. 1).
In this way, one can not only renormalize all single--nucleon Green functions
once and for all, but one also obtains the so--called chiral
logs for all of them. Although one should be wary of doing phenomenology
with chiral logs only, they determine the scale dependence of the renormalized
coupling constants of $O(p^3)$.
The details of the calculation can be found in Ref.~\cite{Ecker}.
A non--trivial part of this calculation consists in finding a heat kernel
representation for the inverse of the differential operator
\beq
iv \cdot \nabla + g_A S \cdot u~.
\eeq
This is precisely the nucleon propagator in the presence
of external fields appearing in the diagrams of Fig. 1.

The renormalization program at $O(p^3)$ can be summarized in the
following way, in complete analogy to the mesonic case
at $O(p^4)$ \cite{GL1,GL2}. By choosing a convenient regularization,
the one--loop functional is decomposed into a divergent and a
finite part. This decomposition introduces an arbitrary scale
parameter $\mu$: although the total one--loop functional is independent
of $\mu$, the two parts are not. The divergent part is then cancelled
by a corresponding piece in the general effective Lagrangian of
$O(p^3)$,
\beq
\cL^{(3)}_{\pi N}(x) = \frac{1}{(4\pi F)^2} \sum_i B_i \ol{N_v}(x)
O_i(x) N_v(x)~, \label{eq:Lct}
\eeq
through the decomposition
\beq
B_i = B_i^r(\mu) + (4\pi)^2 \beta_i \Lambda(\mu) \label{eq:beta}
\eeq
of the dimensionless coupling constants $B_i$. The quantity $\Lambda(\mu)$
is divergent and the coefficients $\beta_i$ are chosen such that
the divergent part of (\ref{eq:Lct}) cancels the divergent
piece of the one--loop functional. The complete generating functional
of $O(p^3)$ then consists of the finite one--loop functional and
the tree--level functional due to (\ref{eq:Lct}),
with the couplings $B_i$ replaced by the renormalized coupling constants
$B_i^r(\mu)$. The complete functional of $O(p^3)$ is finite and independent
of the scale $\mu$ by construction. In Table \ref{tab:beta}, some of
the operators $O_i$ are listed together with their coefficients $\beta_i$.

\begin{table}
\caption{Some counterterms and their $\beta$ functions as defined in
Eqs. (\protect\ref{eq:Lct}), (\protect\ref{eq:beta}). The complete
list of such terms with non--zero $\beta_i$ is given in \protect
\cite{Ecker}.}\label{tab:beta}
$$
\begin{tabular}{|r|c|c|} \hline
i  & $O_i$ & $\beta_i$ \\ \hline
1  & $i[u_\mu,v \cdot \nabla u^\mu]$ & $g^4_A/8$ \\
2  & $i[u_\mu,\nabla^\mu v \cdot u]$ & $- (1 + 5 g^2_A)/12$ \\
3  & $i[v \cdot u, v \cdot \nabla v \cdot u]$ & $(4 - g^4_A)/8$ \\
4  & $S \cdot u \langle u \cdot u \rangle$ & $g_A (4 - g^4_A)/8$ \\ \hline
\end{tabular}
$$
\end{table}

The situation for the pion--nucleon system to $O(p^3)$ is now comparable
to the mesonic sector at $O(p^4)$
\cite{GL1,GL2}. It remains to extract the low--energy constants
$B_i^r(\mu)$, the analogues of the mesonic constants
$L_i^r(\mu)$ \cite{GL2}, as
well as the scale--independent constants of $O(p^2)$ from
pion--nucleon data \cite{BKKM}. Another important task is to try to understand
the actual values of these parameters, in particular to investigate
systematically the effect of meson and baryon resonances.

\renewcommand{\theequation}{\arabic{section}.\arabic{equation}}
\setcounter{equation}{0}
\section{Conclusions}
CHPT is the effective field theory of the standard model in the hadronic
sector at low energies. It is a ``non--renormalizable", yet fully
consistent quantum field theory giving rise to a systematic low--energy
expansion of amplitudes. The relevant scale for this expansion is
$4\pi F_\pi$, which is of the same order of magnitude as the
nucleon mass $m$. The heavy mass expansion for the meson--baryon
part of the effective Lagrangian allows for a simultaneous expansion
in inverse powers of $4\pi F_\pi$ and $m$. The renormalization
has been fully implemented in a manifestly chiral--invariant way to
$O(p^4)$ in the meson and to $O(p^3)$ in the pion--nucleon sector.

Among the future developments in the meson--baryon system are the
renormalization at $O(p^4)$, which again involves only one--loop diagrams,
inclusion of higher baryon states, extension to chiral $SU(3)$
and applications for the non--leptonic weak interactions of baryons.

%\newpage
\newcommand{\PL}[3]{{Phys. Lett.}        {#1} {(19#2)} {#3}}
\newcommand{\PRL}[3]{{Phys. Rev. Lett.} {#1} {(19#2)} {#3}}
\newcommand{\PR}[3]{{Phys. Rev.}        {#1} {(19#2)} {#3}}
\newcommand{\NP}[3]{{Nucl. Phys.}        {#1} {(19#2)} {#3}}

\vfill
\section*{Figure Captions}
\begin{description}
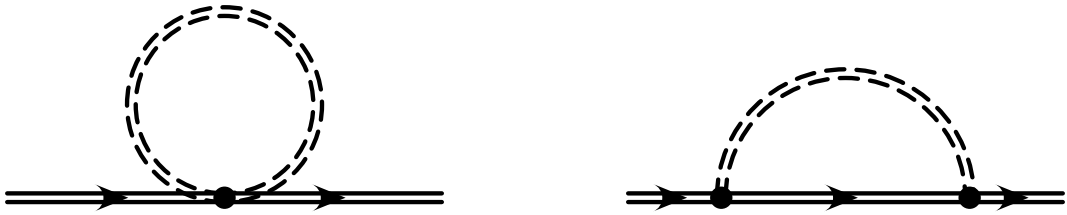
\item[Fig. 1:] Irreducible one--loop diagrams. The full (dashed) lines
denote the nucleon (meson) propagators. The double lines indicate that
the propagators (as well as the vertices) have the full tree--level
structure attached to them as functionals of the external fields.
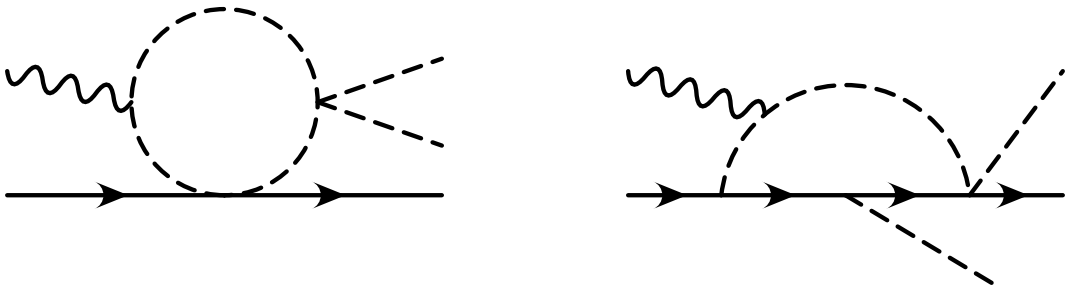
\item[Fig. 2:] Feynman diagrams for  $\pi\pi$ photo--(electro--)production
off nucleons as explicit examples for the diagrams of Fig. 1.
\end{description}

\end{document}